\documentclass[11pt]{article}
\usepackage{moriond,epsfig}

\def\be{\begin{equation}}
\def\ee{\end{equation}}
\def\bea{\begin{eqnarray}}
\def\eea{\end{eqnarray}}

\def\Pomeron{I\hspace{-1.2mm}P}

\begin{document}
\vspace*{4cm}
\title{Exclusive $e^+e^-$, Di-photon and Di-jet Production at the Tevatron}

\author{ K. TERASHI\\
(For the CDF and D\O\hspace{1.4mm}Collaborations)}

\address{The Rockefeller University, 1230 York Avenue, New York, 10021, USA}

\maketitle\abstracts{
Results from studies on exclusive production of electron-position pair, di-photon, and di-jet production at CDF in proton-antiproton collisions at the Fermilab Tevatron are presented. The first observation and cross section measurements of exclusive $e^+e^-$ and di-jet production in hadron-hadron collisions are emphasized.}

\section{Introduction}
\begin{figure}
\begin{center}
\psfig{figure=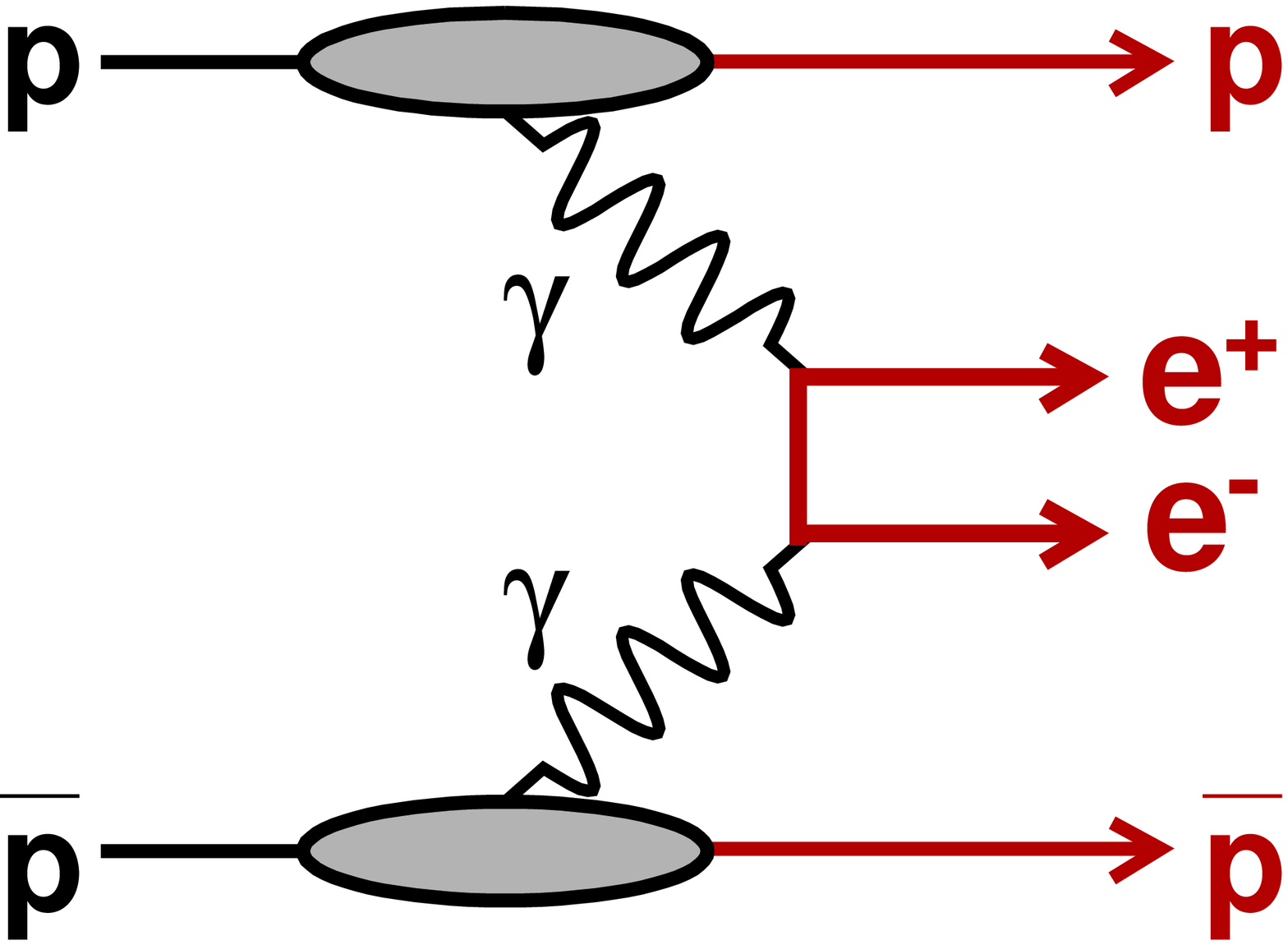,width=3.7cm}\hfill
\psfig{figure=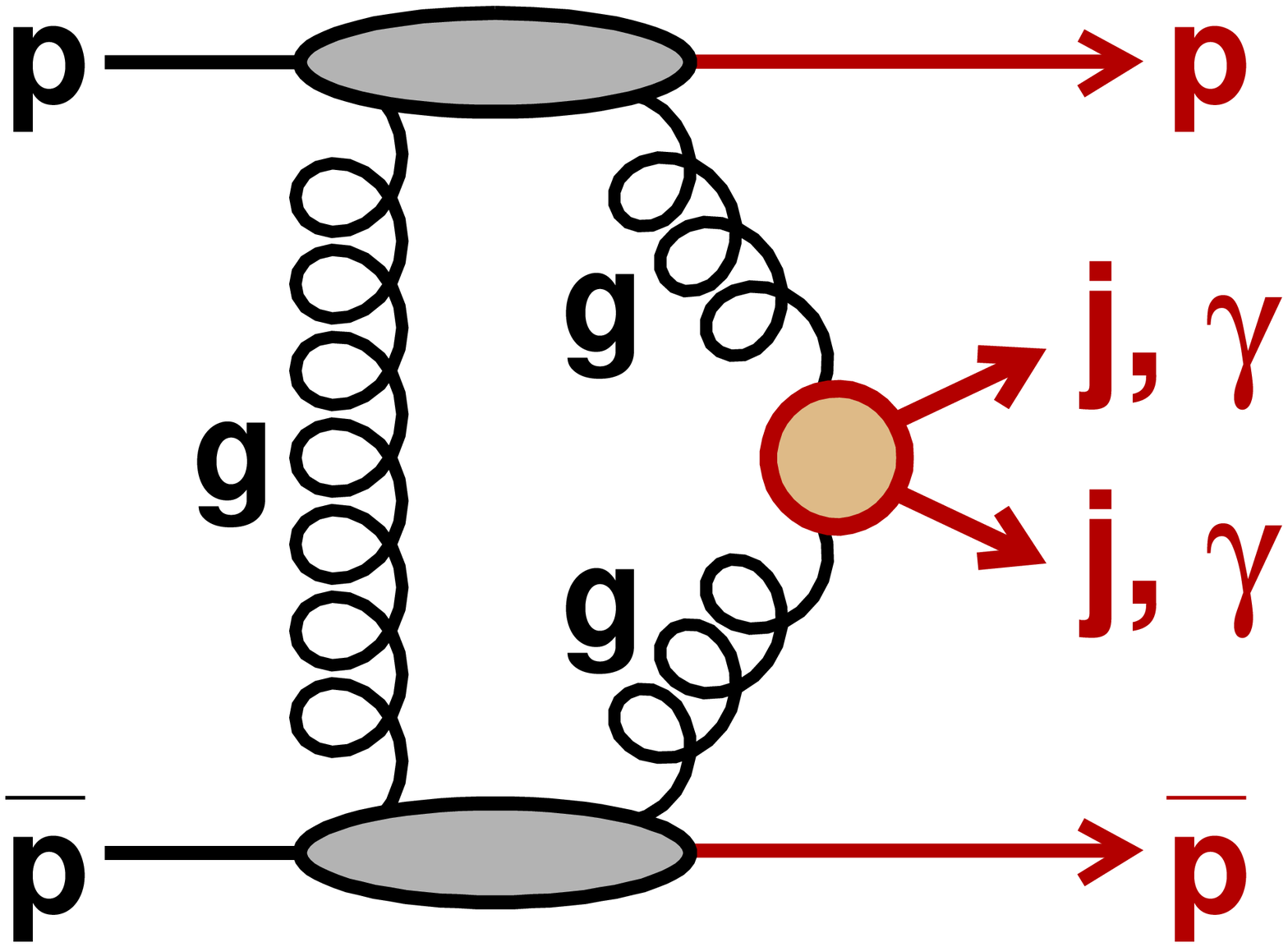,width=3.7cm}\hfill
\psfig{figure=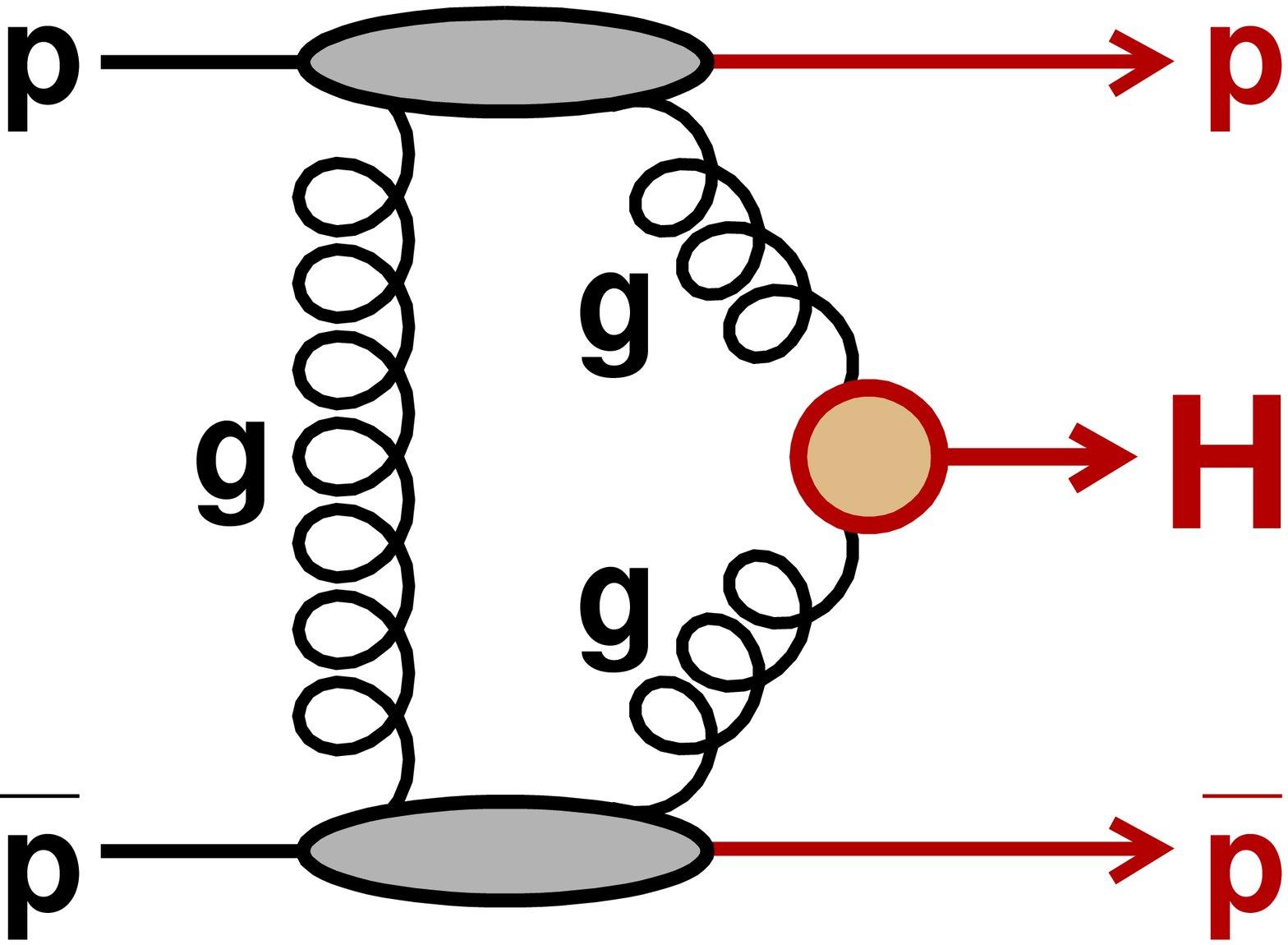,width=3.7cm}
\caption{{\it Left}: Diagram of exclusive $e^+e^-$ production via $\gamma\gamma \rightarrow e^+e^-$, {\it Center} ({\it Right}): Leading order diagrams of exclusive di-jet and di-photon (Higgs boson) production in $\bar{p}p$ collisions.
\label{fig:diagrams}}
\end{center}
\end{figure}

Exclusive production in $\bar{p}p$ collisions is a process in which the proton and antiproton remain intact and an exclusive state $X_{\rm excl}$ of particle(s) is centrally produced: $\bar{p}+p \rightarrow \bar{p}' + X_{\rm excl} + p'$.
Exclusive di-lepton production may occur through a two photon exchange process; $\bar{p}+p \rightarrow \bar{p}'+e^+e^-+p'$ via $\gamma\gamma \rightarrow e^+e^-$, where the lepton is an electron. Since this process, shown in Fig.~\ref{fig:diagrams}~(left), depends essentially only on Quantum Electro-Dynamics, the cross section is known with an accuracy better than 1~\%. 
Exclusive di-jet (or di-photon) production is a special case of di-jet (di-photon) production in double Pomeron exchange (DPE), a diffractive process in which the $p$ and $\bar{p}$ remain intact with a small momentum loss, and a system $X_{\rm incl}$ containing the jets (photons) is produced:
$\bar{p} + p \rightarrow [\bar{p}' + \Pomeron_{\bar p}] + [p' + \Pomeron_p] \rightarrow \bar{p}' + X_{\rm incl} + p'$.
Here,  $\Pomeron$ represents a Pomeron, defined as an exchange consisting of a colorless combination of gluons and/or quarks carrying the quantum numbers of the vacuum. In a particle-like Pomeron picture (e.g. see~\cite{IS}), the system $X_{\rm incl}$ produced by two Pomeron collisions, $\Pomeron_{\bar p} +\Pomeron_p \rightarrow X_{\rm incl} \Rightarrow (jet_1+jet_2)+Y$, generally contains Pomeron remnants $Y$ in addition to jets  (in di-jet case). Di-jet or di-photon production by DPE may occur as an exclusive process with no Pomeron remnants ($Y=0$).
Exclusive di-jet or di-photon production may also occur through a $t$-channel color-singlet two gluon exchange at leading order (LO) perturbative Quantum Chromo-Dynamics (QCD), shown in Fig.~\ref{fig:diagrams}~(center).
A similar diagram (right) is used to calculate exclusive Higgs Boson production~\cite{KMR}. 

In this paper, we report an observation of exclusive $e^+e^-$ production by CDF in hadron-hadron collisions ~\cite{eePRL}. Also, reported is a first observation of exclusive di-jet production in $\bar{p}p$ collisions, and preliminary results on exclusive di-photon search at CDF. 

\section{CDF II Detector}
Among relevant CDF II detectors, crucial for detecting the rapidity gap (defined as a region of pseudorapidity devoid of particles) in the exclusive events are the forward detectors which consist of the MiniPlug calorimeters (MPCAL)~\cite{MPCAL}, the Beam Shower Counters (BSC), a Roman Pot Spectrometer (RPS), and Cerenkov Luminosity Counters (CLC)~\cite{CLC}.  The MPCAL, designed to measure particle energies in the region $3.6<|\eta|<5.2$, consist of alternating lead plates and liquid scintillator layers perpendicular to the beam, which are read out by wave-length shifting fibers. Combining with the central (CCAL) and plug (PCAL) calorimeters, the MPCAL extends the calorimeter coverage up to $|\eta|=5.2$.
The BSC are scintillation counters surrounding the beam pipe at three (four) different locations on the outgoing $p$ ($\bar{p}$) side of the CDF~II detector. Covering the range $5.4<|\eta|<5.9$ is the BSC1 system, which is used for triggering on events with forward rapidity gaps. The RPS, located at $\sim57$ m downstream in the $\bar{p}$ beam direction, consists of three Roman pot stations, each containing a scintillation counter used for triggering on the $\bar{p}$, and a scintillation fiber tracking detector for measuring the position and angle of the detected $\bar{p}$. The CLC covering the range $3.7<|\eta|<4.7$ are used to measure the luminosity and, in some cases, to sharpen the rapidity gap in the MPCAL.

\section{Data Samples}\label{sec:data}
Main data samples used in the studies presented in this paper are collected with two dedicated diffractive triggers. Search for exclusive $e^+e^-$ and di-photon production uses the data triggered on two clusters of energy in the CCAL or PCAL calorimeters, and no activity in the BSC1 counters on both sides.  Off-line, events containing electron or photon candidates with $E_T>5$~GeV are selected. The total integrated luminosity of this sample is $532\pm32$ pb$^{-1}$.

Exclusive di-jet search is performed on a data sample triggered when the event contains 1) a 3-fold coincidence of the RPS trigger counters in time with a $\bar{p}$ gate, b) at least one trigger tower with $E_T>5$~GeV, and c)  no BSC1 activity on the outgoing $p$-side (referred to as RP+GAP+Jet5 trigger). The events in the sample with an integrated luminosity of $313\pm19$ pb$^{-1}$ are required to pass the following selection cuts; 1) no more than one reconstructed primary vertex within $|z|<60$~cm, 2) RPS trigger counter pulse height cut, 3) at least two jets (corrected to hadron level) with $E_T>10$~GeV within $|\eta|<2.5$, 4) $0.01<\xi_{\bar{p}}^X<0.12$, and 5) zero MPCAL and zero CLC hit multiplicity on the $p$ side.
Cut 1) is used to reduce overlap events with multiple $\bar{p}p$ collisions occurring during the same bunch crossing. Cut 2) is imposed to reject triggers caused by particles hitting the beam pipe in the vicinity of the RPS. The $\xi_{\bar{p}}^X$, defined as $\xi_{\bar{p}}^X=\sum_{i=1}^{N_{tower}}(E_T^i\,e^{-\eta^i})/\sqrt{s}$ where the sum is carried over all calorimeter towers with $E_T>100$~MeV for CCAL and PCAL, and $E_T>20$~MeV for MPCAL, representing the fractional momentum loss of the $\bar{p}$, is required to be in the range of cut 4) to reduce remaining overlap events. Finally, the DPE contribution in the sample is enhanced by the cut 5) which requires a wide $p$-side rapidity gap in the region of $3.6<\eta<5.9$. 

\section{Exclusive Electron-Positron Production}
Search for exclusive $e^+e^-$ signal is based on selecting events with electron and position candidates and refining the selected sample by requiring no additional particle signatures in the calorimeters or BSC. The electron (or position) candidate is defined as a cluster in the EM calorimeter with $E_T>5$~GeV and $|\eta|<2$, being consistent with an $e^-$ ($e^+$), and a track with $p_T>1$~GeV/c pointing to the calorimeter cluster. A particle signature in the calorimeter (BSC) is defined as a cluster of adjacent towers in the MPCAL or a single tower in the CCAL or PCAL (any BSC hit) above the noise threshold. 
From the sample of triggered events,  a total of 16  candidate events pass electron identification, cosmic ray rejection, and the veto on additional particle signatures.

The total background, which consists of a) jets faking electrons, b) cosmic rays interacting in the detector, c) non-exclusive events, and b) $\gamma\gamma \rightarrow e^+e^-$ events with
proton dissociation, is estimated to be $1.9\pm0.3$ events in the 16 candidate events. The dominant contribution is proton dissociation events ($1.6\pm0.3$ events). 
The cross section for exclusive $\bar{p}p \rightarrow \bar{p}+e^+e^-+p$ is measured to be $1.6^{+0.5}_{-0.3}({\rm stat})\pm0.3({\rm syst})$~pb, which is in good agreement with the theoretical cross section of $1.71\pm0.01$~pb obtained from {\sc LPAIR} Monte Carlo (MC) simulation~\cite{LPAIR}.  The probability of observing 16 events or more when $1.9\pm0.3$ events are expected is $1.3\times10^{-9}$, equivalent to a $5.5\sigma$ effect. The good agreement between the measured and expected cross sections proves that rapidity gap signature in the detectors is well understood. 

\section{Exclusive Di-photon Production}
Search for exclusive di-photon event, $\bar{p}p \rightarrow \bar{p}+\gamma\gamma+p$,  follows the same event selections as the exclusive $e^+e^-$  search, except that the photon candidates (defined as EM calorimeter clusters with $E_T>5$~GeV and $|\eta|<1$) have no tracks pointing to the clusters. The total event selection efficiency for exclusive $\gamma\gamma$ events with photons of $E_T>5$~GeV and $|\eta|<1$ is $4.0\pm0.7$~\%.  

In the same sample as used in exclusive $e^+e^-$ analysis, three events  passed the selection criteria. 
The ExHuME MC generator~\cite{ExHuME} for exclusive di-photon production via $gg \rightarrow \gamma\gamma$ processes predicts $1^{+3}_{-1}$ events, which is consistent with the number of observed candidate events. Background estimation is currently under way.

\section{Exclusive Di-jet Production}
A special analysis strategy is developed to search for the exclusive di-jet production as it is quite difficult to identify particles associated with the jets. The analysis employs distributions of ``di-jet mass fraction'', $R_{jj}$, defined as the di-jet invariant mass $M_{jj}$ divided by the whole system mass $M_X$ (except leading nucleons); $R_{jj}=M_{jj}/M_X$. The $M_X$ is obtained from all 
calorimeter towers above the thresholds used to calculate the $\xi_{\bar{p}}^X$, and the $M_{jj}$ is calculated from calorimeter tower energies inside the R=0.7 cones of jets.
The exclusive signal is extracted by comparing the DPE di-jet data selected in Sec.~\ref{sec:data} with inclusive DPE di-jet events (which do not contain exclusive di-jet production) in the $R_{jj}$ distribution shape. We use POMWIG MC generator~\cite{POMWIG} with detector simulation to simulate the DPE di-jets. 

\begin{figure}
\begin{center}
\psfig{figure=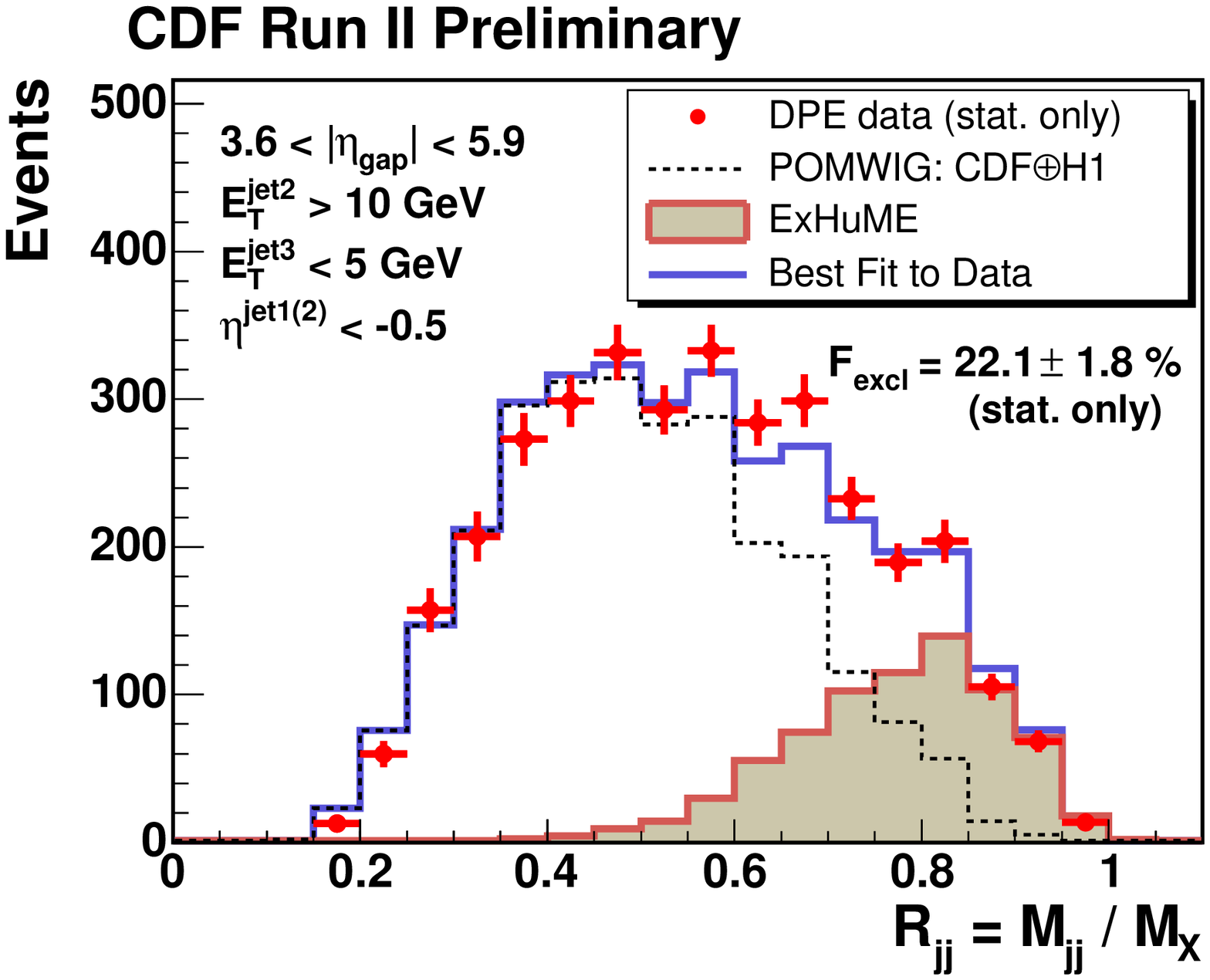,width=5.1cm}\hfill
\psfig{figure=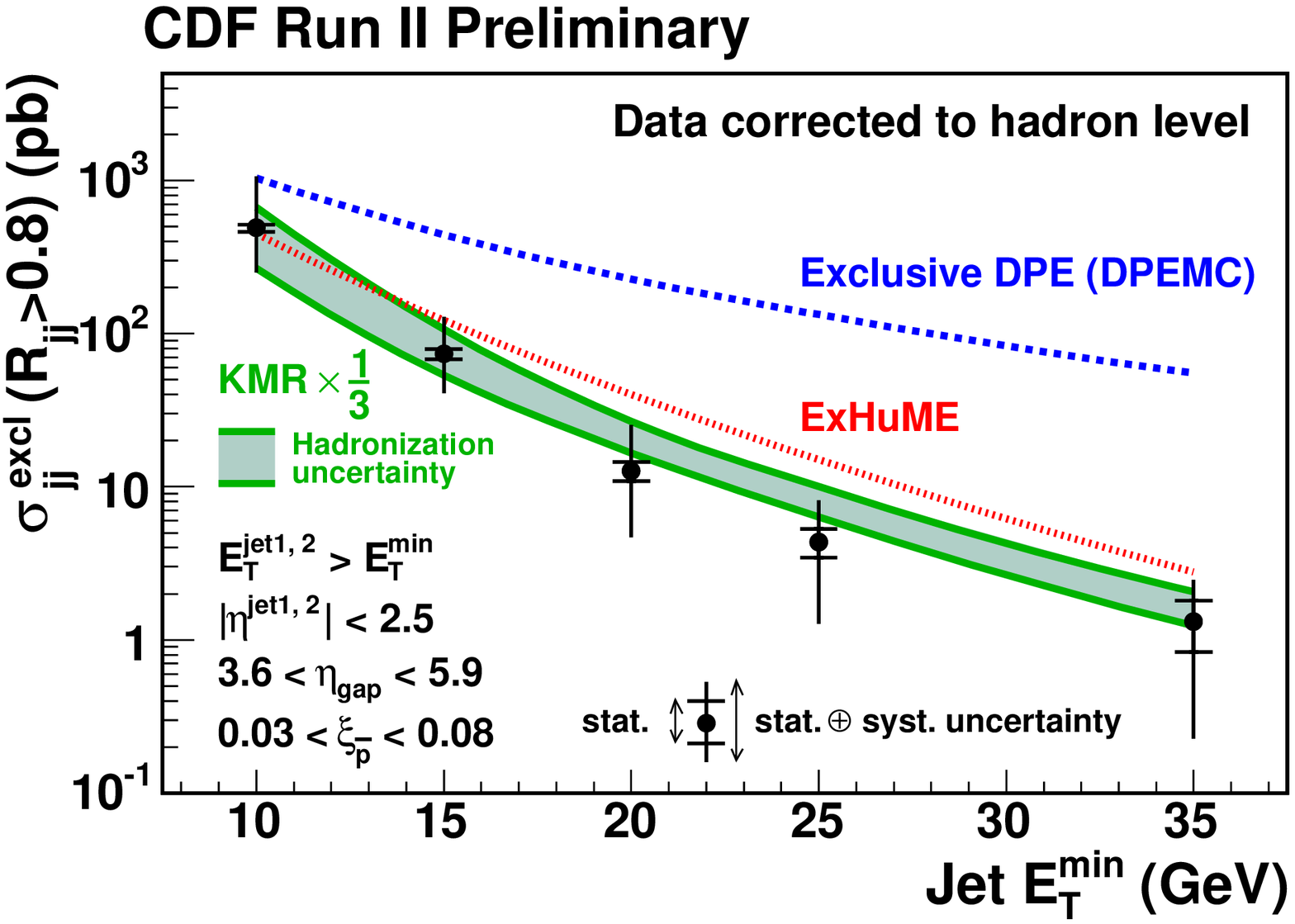,width=5.25cm}\hfill
\psfig{figure=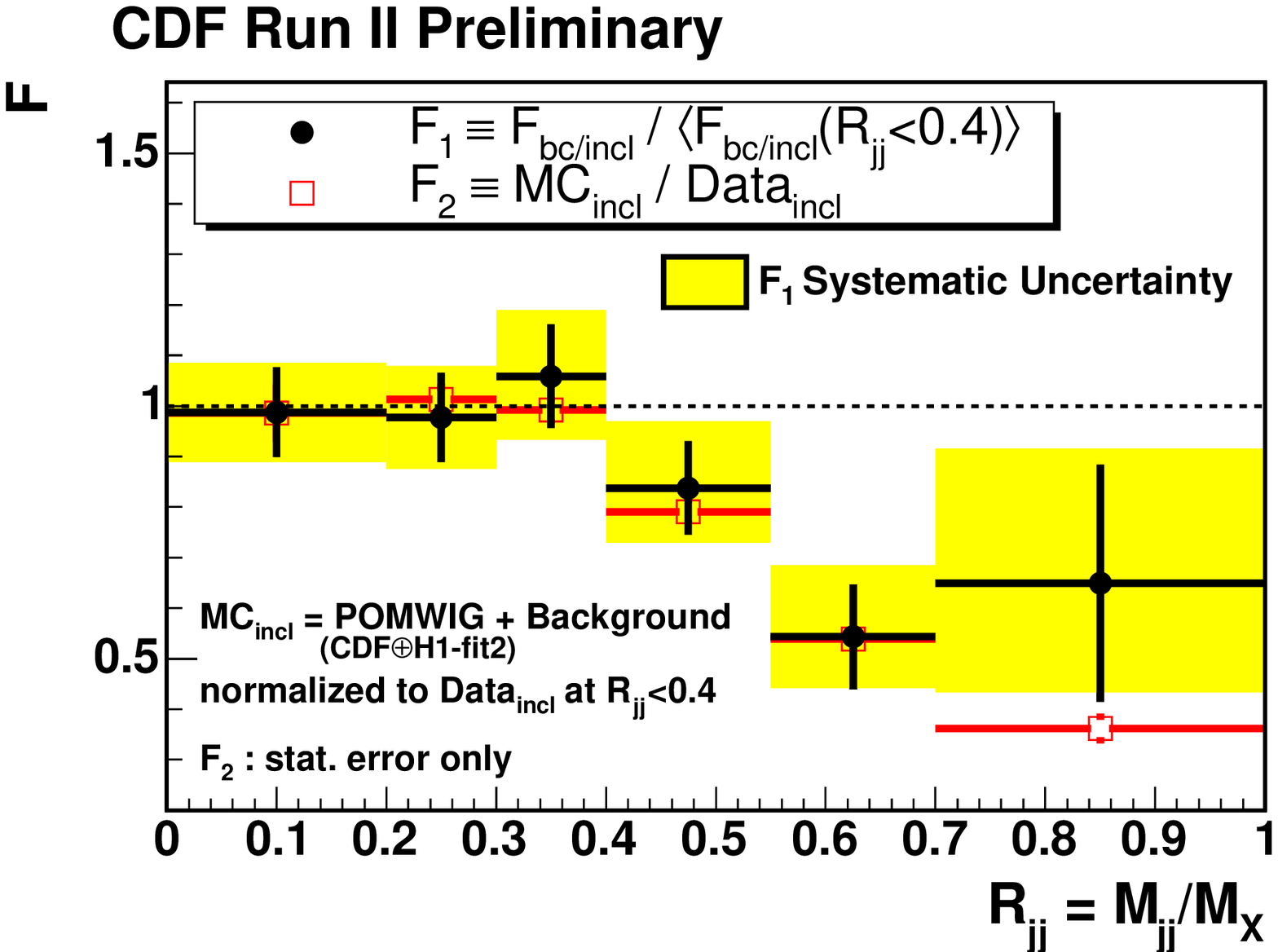,width=5.25cm}
\caption{{\it Left}: Dijet mass fraction in DPE data (points) and best fit (solid histogram) to the data obtained from {\sc POMWIG} MC events (dashed histogram) and {\sc ExHuME} di-jet MC events (shaded histogram). {\it Center}: Measured exclusive di-jet cross section for $R_{jj}>0.8$ (points) as a function of minimum second jet $E_T$, compared with predictions from {\sc ExclDPE} (dashed curve) and {\sc ExHuME} (dotted curve) MC simulations, and the KMR calculation at LO parton level, scaled down by a factor 3 (shaded band). {\it Right}:  Values of $F_1$ (filled points) and $F_2$ (open points) as a function of $R_{jj}$, where $F_1$ is the ratio of heavy flavor jets to all inclusive jets, normalized to the weighted average value in the region of $R_{jj}<0.4$ (systematic uncertainties are shown by the shaded band), and $F_2$ is the ratio of {\sc POMWIG} MC to inclusive DPE di-jet data.
\label{fig:excl_jj}}
\end{center}
\end{figure}

The $R_{jj}$ distribution comparisons showed a clear excess of data at high $R_{jj}$, as expected for exclusive dijet signal. It turns out that the observed excess is robust against the choice of Pomeron parton distribution functions and Pomeron remnant effects on underlying events. In order to extract the signal from the data, two exclusive di-jet MC simulations, ExHuME~\cite{ExHuME} and exclusive DPE model ({\sc ExclDPE}) in DPEMC~\cite{DPEMC} have been studied. Fig.~\ref{fig:excl_jj}~(left) shows the fit to the data $R_{jj}$ shape using a combination of POMWIG inclusive DPE di-jets and ExHuME exclusive di-jets for events with di-jets of $E_T>10$ GeV and the third jet veto of $E_T^{jet3}<5$ GeV. The third jet veto is introduced because the exclusive MC generates only LO $gg \rightarrow gg$ process. The fit shows the data excess is well described by the presence of exclusive di-jets. The {\sc ExclDPE} provides similar results in the shape fit. From the MC fits to the data, we measure the cross section of exclusive di-jet production as a function of minimum second jet $E_T$, as shown in Fig.~\ref{fig:excl_jj}~(center). The data prefer {\sc ExHuME} and perturbative QCD calculations at LO parton level (KMR) in Ref.~\cite{KMR}.

Exclusive $gg \rightarrow q\bar{q}$ production is expected to be suppressed at LO by the $J_z=0$ total angular momentum selection rule. Confirming the suppression will therefore support the results obtained from the MC based method described above. Since the suppression holds for any quark flavor, we measure the ratio $F_1$ of heavy flavor ($b$ or $c$) quark jets to all jets as a function of $R_{jj}$ using a 200 pb$^{-1}$ sample of data triggered on RP+GAP+Jet5 and a transversely displaced track of $p_T>2$~GeV.
The result, presented in Fig.~\ref{fig:excl_jj}~(right), shows $F_1$ versus $R_{jj}$, where the $F_1$ is normalized by the weighted average value in the range of $R_{jj}<0.4$. The decreasing trend of $F_1$ with $R_{jj}$, which indicates a manifestation of the $J_Z=0$ selection rule, is compared with the MC based result given as $F_2$, where $F_2$ is the ratio of the inclusive MC predicted events (normalized to the data at $R_{jj}<0.4$) to the data. The two results are consistent with each other in both magnitude and $R_{jj}$ dependence.



\section*{References}
\begin{thebibliography}{99}
  \bibitem{IS}G. Ingelman and P. Schlein, Phys. Lett. B {\bf 152}, 256 (1985).
  \bibitem{KMR}
  V. A. Khoze, A. D. Martin, and M. G. Ryskin, Eur. Phys. J. C {\bf 34}, 327 (2004), and references there in; A. B. Kaidalov {\it et al.}, Eur. Phys. J. C {\bf 31}, 387 (2003); {\bf 33}, 261 (2004).
  \bibitem{eePRL}A. Abulencia {\it et al.} (CDF Collaboration), Phys. Rev. Lett. {\bf 98}, 112001 (2007).
  \bibitem{MPCAL}M. Gallinaro {\it et al.}, Nucl. Instrum. Methods A {\bf 496}, 333 (2003).
  \bibitem{CLC}D. Acosta {\it et al.}, Nucl. Instrum. Methods A {\bf 494}, 57 (2002).
  \bibitem{LPAIR}J. Vermaseren, Nucl. Phys. {\bf B229}, 347 (1983).  
  \bibitem{ExHuME}J. Monk and A. Pilkington, Comput. Phys. Commun. {\bf 175}, 232 (2006).
  \bibitem{POMWIG}B. E. Cox and J. R. Forshaw, Comput. Phys. Commun. {\bf 144}, 104 (2002).
  \bibitem{DPEMC}M. Boonekamp and T. Kucs, Comput. Phys. Commun. {\bf 167}, 217 (2005).
\end{thebibliography}
\end{document}